\documentclass[10pt]{article}
\usepackage{times}

\begin{document}

\title{(Quantum) Chaos in BECs}
\author{S.A. Gardiner}

\maketitle

The achievement of Bose-Einstein condensation of a dilute vapour of Rubidium atoms
in 1995 \cite{cornell} heralded the beginning of a rapidly growing field of experimental and
theoretical endeavour, each of which has further stimulated the other. 
In the intervening years, the experimental production of Bose-Einstein
condensates (BECs), particularly of Rubidium and Sodium vapours, has become 
almost routine, and attention has shifted to either the condensation of more 
exotic species (of particluar note is the recent successful condensation of metastable
helium \cite{helium}); or to deeper experimental investigations of the properties of
an atomic BEC, up to the point of viewing a BEC more as a tool rather than an 
end in itself. 
The purpose here is to show links between chaotic dynamics, both in a classical
and a quantum mechanical sense, and the possibly ``chaotic'' 
dynamical and stability properties of BECs.

To begin:
considering only the dynamics of a single {\em classical\/} 
point particle of mass $m$, 
one can reasonably speak of a class of ``chaos-inducing'' 
potentials $V_{c}({\mathbf{x}})$, 
such that a particle can exhibit chaotic dynamics under the influence of such a
potential. 
One can qualitatively visualize chaotic dynamics as being highly
irregular motion in {\em phase space}. For a particle moving in one spatial
dimension this is a two-dimensional space, where (subject to canonical changes
of variables) the axes correspond to the particle's position $x$ and momentum $p$. 
Such highly irregular motion is generally associated with {\em exponential
sensitivity to initial conditions}, i.e.\ small differences in a point
particle's initial position and momentum can make a large difference in its
long-time behaviour. The appropriate classical (Hamilton's) equations of motion 
can then be derived from a Hamiltonian function of the general form:
\begin{equation}
H = %\sum_{k} \frac{p_{k}^{2}}{2m}
\frac{{\mathbf{p}}^{2}}{2m}
 + V_{c}({\mathbf{x}}).
\end{equation}
One can readily make the jump to quantum mechanics by replacing the number-valued 
position and momentum variables with {\em operator} quantities. For a single
particle, the dynamics are then generally most conveniently described by the
following {\em Schr\"{o}dinger equation}:
\begin{equation}
i\hbar\frac{\partial \psi({\mathbf{x}})}{\partial t} 
= \left[
-\frac{\hbar^{2}\nabla^{2}}{2m} + V_{c}({\mathbf{x}})
\right]\psi({\mathbf{x}}).
\end{equation}
This is clearly a {\em linear\/} partial differential equation, where the time-derivative of the
wavefunction $\psi({\mathbf{x}})$ is proportional to the Hamiltonian {\em operator\/} 
[the momenta ${\mathbf{p}}=(p_{x},p_{y},p_{z})$ have been replaced by 
differential operators] acting on
$\psi({\mathbf{x}})$. As chaos is generally associated with nonlinear
differential equations, this can cause some to question how one can speak of
quantum chaos at all. In classical mechanics we are
interested in the motion of the particle in phase space, and thus in the changes
of its position and momentum with time. The classical dynamics of a point
particle's position and momentum, and the dynamics of the quantum wavefunction
$\psi({\mathbf{x}})$ are thus not really directly comparable: better analogies exist
between Heisenberg's equations and Hamilton's equations of motion, or
alternatively between Schr\"{o}dinger's equation and Liouville's equation. 
Here quantum chaos is considered to be the study of quantum dynamical systems, 
the classical
limits of which are capable of exhibiting chaotic dynamics. 
That Schr\"{o}dinger's equation is a linear partial differential equation is 
in this context an
irrelevancy; nevertheless it is interesting to consider the effect of
introducing an explicit nonlinearity into a quantum-chaotic Schr\"{o}dinger 
equation:
\begin{equation}
i\hbar\frac{\partial \psi({\mathbf{x}})}{\partial t} = \left[
-\frac{\hbar^{2}\nabla^{2}}{2m} + V_{c}({\mathbf{x}})
+u|\psi({\mathbf{x}})|^{2}
\right]\psi({\mathbf{x}}).
\end{equation}
One could regard such an equation as describing 
``nonlinear quantum chaos'' or ``nonlinear wave chaos.'' Such {\em nonlinear
Schr\"{o}dinger equations\/} are relevant for nonlinear optics, but more
pertinently are extensively used to describe 
BECs, when they are called
{\em Gross-Pitaevskii\/} equations \cite{shlyapnikov}. 
A Gross-Pitaevskii equation is used as an
{\em approximate\/} description of the dynamics of a large number $N$ of
(bosonic) interacting quantum mechanical particles, which is more
 fully described by a quantum
field; if nearly all of these particles can be
considered to be in the same motional state $\psi({\mathbf{x}})$ 
(i.e.\ if we have a BEC), 
then it is possible to describe
most of the underlying quantum field by a classical field, and to regard what is
left over as being ``small.'' The Gross-Pitaevskii equation is then the equation
of motion of this classical field.
It is nevertheless sometimes necessary to consider the dynamics of the full
quantum field; in second-quantized form, the Hamiltonian operator leading to
such a Gross-Pitaevskii equation (with $u=gN$ and $g$ being a parameter
essentially describing how strongly individual particles interact) is given by
\begin{equation}
\hat{H} = \int d^{3}{\mathbf{x}} \hat{\psi}^{\dagger}({\mathbf{x}})\left[
 -\frac{\hbar^{2}\nabla^{2}}{2m} + V_{c}({\mathbf{x}})
+\frac{g}{2}\hat{\psi}^{\dagger}({\mathbf{x}})\hat{\psi}({\mathbf{x}})\right]
\hat{\psi}({\mathbf{x}}),
\end{equation}
where $\hat{\psi}^{\dagger}({\mathbf{x}})$ and 
$\hat{\psi}({\mathbf{x}})$ are bosonic field {\em operators}, which create and
annihilate, respectively, a particle at position ${\mathbf{x}}$. In this context,
one could thus possibly speak of
``quantum field chaos'' when considering the dynamics induced by such a
Hamiltonian.

A brief discussion of integrability, a concept closely tied to the (non)
chaoticity of dynamical systems, is now necessary \cite{kus}. Recalling the
classical dynamics of a {\em single\/} particle, such a system is considered
integrable if
there exist as many independent conserved quantities as there are motional
degrees of freedom (i.e.\ the number of relevant spatial dimensions). 
If the system {\em is\/} integrable, then the phase-space dynamics of the 
particle are so restricted by the necessity of observing these
conservation laws that chaos is impossible. 
To take the rather trivial example of the one-dimensional harmonic oscillator
[potential $V(x) = m\omega^{2}x^{2}/2$], the solution of Hamilton's equations of
motion (in slightly unconventional form) is given by:
\begin{equation}
\alpha(t) = e^{-i\omega t}\alpha(0),
\label{harmonic}
\end{equation}
where $\alpha$ is a complex variable, defined as
\begin{equation}
\alpha = \frac{m\omega x + i p}{\sqrt{2m}}.
\end{equation}
It is clear that the particle motion is restricted to closed circles 
in the two-dimensional phase space, making both ``irregular'' motion in phase
space, and sensitivity to initial conditions completely impossible.
In the language of Hamilton-Jacobi theory, $|\alpha(t)|$ is an
action variable, which is conserved (and equal to the square root of the
energy), and the dynamics are fully described by the time-evolution of 
the phase angle of $\alpha(t)$, which is then the canonically conjugate 
angle variable $\theta(t)$ to $|\alpha(t)|$, defined by
\begin{equation}
\theta = \arctan\left(\frac{p}{m\omega x}\right).
\end{equation}
Similar analyses can be made for any one-dimensional, time independent system;
due to the time-independence of the Hamiltonian, the energy is always a
conserved quantity, which is enough to make a one-dimensional system integrable.
Two- and three-dimensional systems will require additional {\em independent\/}
conserved quantities to be integrable, and time dependence in a one-dimensional
system generally lifts the property of integrability.
Concepts of quantum integrability generally, by analogy, consider conserved
{\em observable\/} quantities, which in quantum mechanics are described by
operators \cite{kus}; it is possible to consider something different 
however. 

A wavefunction $\psi({\mathbf{x}})$ can generally be decomposed in terms of some
complete orthonormal basis \cite{discrete}:
\begin{equation}
\psi({\mathbf{x}}) = \sum_{k=0}^{\infty}c_{k}\varphi_{k}({\mathbf{x}}).
\end{equation}
If additionally the $\varphi_{k}(x)$ are eigenfunctions of the (assumed 
time-independent) Hamiltonian, then any time dependence of $\psi(x)$ is
contained completely within the coefficients $c_{k}$: thus
\begin{equation}
c_{k}(t) = e^{-i\omega_{k}t}c_{k}(0).
\label{coefficient}
\end{equation}
For example, in the case of a one-dimensional {\em quantum\/} harmonic
oscillator, the $\varphi_{k}(x)$ are Gauss-Hermite polynomials, and
$\omega_{k}=\omega(k+1/2)$. 
There is an now an obvious similarity between equations (\ref{harmonic}) and
(\ref{coefficient}). Extending the analogy, we can consider motion in an
infinite-dimensional pseudo phase space of a ``point particle'' having 
``position'' values 
\begin{equation}
\chi_{k}=
c_{k}+c_{k}^{*},
\end{equation}
and ``momentum'' values of 
\begin{equation}
\rho_{k} =
-i(c_{k}-c_{k}^{*}).
\end{equation}
Within this pseudo phase space however, as in the case of the classical harmonic
oscillator, motion is severely restricted by the fact that all the $|c_{k}|$ are
conserved \cite{norm}. The $|c_{k}|$ (effectively the ``action variables'' in the
pseudo phase space) in fact form a {\em complete set\/} of conserved
quantities (i.e.\ there are as many conserved quantities as degrees of freedom), 
meaning that, in this sense, the Schr\"{o}dinger equation is {\em
always\/} integrable and thus {\em never\/} chaotic, a fact which is indeed a
direct consequence of the Schr\"{o}dinger equation's linearity. 

For a general Gross-Pitaevskii equation, the additional nonlinear term means that
integrability clearly cannot be taken for granted. It is, for example, known that
the one dimensional, homogeneous case
\begin{equation}
i\hbar \frac{\partial \psi(x)}{\partial t} = \left[
-\frac{\hbar^{2}}{2m}\frac{\partial^{2}}{\partial x^{2}} + u|\psi(x)|^{2}
\right]\psi(x)
\end{equation}
is integrable \cite{nonlinear}, and that the conserved quantities can be determined using inverse
scattering techniques. However, even for the seemingly trivial extension of simply
adding a harmonic potential (generally more relevant to current experiments), it
is unknown whether the resulting Gross-Pitaevskii equation is integrable or not;
the application of a $\delta$-kicking potential (as is widely considered 
in the study of classical and quantum chaos)
\begin{equation}
V_{c}(x)=K\cos(kx)\sum_{n=-\infty}^{\infty}\delta(t-nT),
\end{equation}
will produce a Gross-Pitaevskii equation which is unlikely to be integrable
\cite{me}.
This general lack of integrability therefore raises the possibility of chaotic 
dynamics in the pseudo phase space, including extreme sensitivity to initial
conditions.

In classical machanics, sensitivity to initial conditions is generally gauged by
the Lyapunov exponent, so that a positive Lyapunov is considered a strong
indicator of chaotic dynamics. The Lyapunov exponent is concerned with the growth
with time
of the euclidean ``distance'' $d$ 
in phase space between two initially close point particles. In one dimension, $d$
is thus given by
\begin{equation}
d(t)= \sqrt{[x_{1}(t)-x_{2}(t)]^{2}+[p_{1}(t)-p_{2}(t)]^{2}}.
\end{equation}
In the pseudo phase space produced by some orthonormal representation of two
initial wavefunctions $\psi_{1}(x)$ and $\psi_{2}(x)$, 
the equivalent $d$ is given by
\begin{eqnarray}
d(t) &=& 
\sqrt{\sum_{k}[\chi_{k1}(t)-\chi_{k2}(t)]^{2}+[\rho_{k1}(t)-\rho_{k2}(t)]^{2}}
\nonumber \\ 
&=& \sqrt{2-
\int d^{3}{\mathbf{x}} \left[\psi_{1}^{*}({\mathbf{x}},t)\psi_{2}({\mathbf{x}},t) + 
\psi_{2}^{*}({\mathbf{x}},t)\psi_{1}({\mathbf{x}},t)
\right]},
\end{eqnarray}
where in the case of the linear Schr\"{o}dinger equation, by unitarity 
$\int d^{3}{\mathbf{x}} \psi_{1}^{*}({\mathbf{x}})\psi_{2}({\mathbf{x}})$ 
cannot change, and $d$ will therefore be
constant \cite{phase}. Upon the addition of a nonlinearity this is of course no longer
guaranteed, which in the case of the Gross-Pitaevskii equation
can have important consequences for the stability of the condensate, due to
a correspondance between the propagation of linearized perturbations
around the condensate wavefuntion, and propagation (and growth) of the
non-condensate fraction.
 
It is now necessary to return briefly to a fuller quantum mechanical description of
the BEC. If the number of particles in the condensate mode is
nearly $N$, the field operator $\hat{\psi}({\mathbf{x}})$ can be written as
\cite{gardiner,castin}
\begin{equation}
\hat{\psi}({\mathbf{x}})\approx 
\hat{a}
\left[
\psi
({\mathbf{x}}) + \frac{1}{\sqrt{%\hat{
N%}
}}\sum_{k}\hat{b}_{k}u_{k}({\mathbf{x}}) +
\hat{b}_{k}^{\dagger}v_{k}^{*}({\mathbf{x}})
\right].
\end{equation}
The $\hat{a}$ operator annihilates a particle in the condensate mode
$\psi({\mathbf{x}})$ (it
is approximately this mode which is propagated by the Gross-Pitaevskii equation if
the number of particles in the condensate mode $\hat{a}^{\dagger}\hat{a}\approx N$),
and the $\hat{b}_{k}^{\dagger}$ and $\hat{b}_{k}$ are creation and annihilation
operators of {\em excitations\/} above the condensate, the $u_{k}({\mathbf{x}})$ and
$v_{k}^{*}({\mathbf{x}})$ describing the spatial dependence of these excitations. These
excitation modes clearly describe {\em non-condensate\/} particles, and for the
idealized case of zero temperature, the number of non-condensate particles is
then given by 
\begin{equation}
\sum_{k}\int d^{3}{\mathbf{x}} |v_{k}({\mathbf{x}},t)|^{2}.
\end{equation}
The excitation modes are propagated by slightly modified Bogoliubov
equations, which are identical to the equations derived by considering
perturbations around and {\em orthogonal\/} to $\psi({\mathbf{x}})$ propagated by the
Gross-Pitaevskii equation up to {\em linear\/} order in the perturbation only.
If one then considers
\begin{equation}
\psi_{1}({\mathbf{x}},0) = \psi({\mathbf{x}},0),
\end{equation} and 
\begin{equation}
\psi_{2}({\mathbf{x}},0)= \psi({\mathbf{x}},0)
+\frac{1}{\sqrt{N}}[u_{k}({\mathbf{x}},0) +
v_{k}^{*}({\mathbf{x}},0)], 
\end{equation}
it is clear that a reduction with time of $|\int d^{3}{\mathbf{x}}
\psi_{1}^{*}({\mathbf{x}},t)\psi_{2}({\mathbf{x}},t)|$ directly implies growth in the number of 
non-condensate particles, and, more generally, growth in the number of
non-condensate particles is implied by any kind of linear instability in a
solution of the Gross-Pitaevskii equation.
Chaotic dynamics in 
the pseudo phase space thus
imply the rapid increase in number of
non-condensate particles, which in turn implies rapid depletion of the
condensate. There is a caveat to this: 
the derivation of the equations used in the analysis
formally take the
limit $N \rightarrow \infty$, $g\rightarrow 0$, while keeping $gN$ constant. There
is therefore {\em formally\/} an infinite number of condensate particles, which stays infinite as
the number of non-condensate particles increases, and the Gross-Pitaevskii
equation is always valid.

In any real situation the total number of particles must of course be finite. 
Recalling that it is assumed that nearly all the particles are in the condensate
mode $\psi({\mathbf{x}})$, it is clear that once
depletion starts to become significant, this approach can be problematical. 
In the systematic (and equivalent) expansions of C.W. Gardiner \cite{gardiner}, 
and of Castin and
Dum \cite{castin}, there are in principle further terms which can be incorporated to account for
higher order effects. Morgan \cite{morgan} has taken a slightly different approach by only
requiring that the occupation of the condensate mode be large, i.e.\ it may be
significantly different from the total particle number, and adding in
perturbative corrections to the Gross-Pitaevskii and Bogoliubov equations. This
treatment, however, is designed for studying essentially steady-state BECs at finite temperature, rather than the explicitly dynamical sorts of 
situations described here. %It may even be possible in some cases to solve the 
%dynamics of the entire $N$ body system numerically. 

The {\em theoretical\/} treatment of such situations is clearly a difficult problem, it is however 
likely relevant to coming generations of BEC experiments. If one actually wants 
to {\em do\/} something with a BEC, this will involve
dynamics. As generic Gross-Pitaevskii equations are presumably non-integrable, 
there exists the definite possibility of chaos and instability in some 
circumstances; this in turn implies the possibility of rapid depletion of the
condensate mode, and concomitant loss of the desirable property of 
{\em coherence\/} of the  macroscopic matter wave. 
Given this, it is equally desirable that such processes be better understood;
this thus seems a suitable ``quantum challenge'' for the 21st century!


\begin{thebibliography}{99}
\bibitem{discrete}
For simplicity a discrete decomposition has been assumed. This restriction is
unlikely to be qualitatively important however.
\bibitem{norm}
The additional requirement that the wavefunction be normalized to 1 is essentially
equivalent to a restriction on the set of possible initial conditions.
\bibitem{phase}
A word of caution: generally the {\em global phase\/} of a wavefunction is not
considered to have any physical meaning; in the pseudo phase space described
here, however, this is not accounted for.
\end{thebibliography}
\end{document}